\documentclass{elsart}
\usepackage{graphicx}  
\usepackage{amssymb}
\usepackage{rotating}
\graphicspath{{fig/}}
\begin{document}

\begin{frontmatter}
\title{Preparation of Neutron-activated Xenon for Liquid Xenon Detector Calibration}

\author{K.~Ni,}
\author{R.~Hasty,}
\author{T.M.~Wongjirad,}
\author{L.~Kastens,}
\author{A.~Manzur,}
\author{D.N.~McKinsey\corauthref{cor1}}
\ead{daniel.mckinsey@yale.edu}
\corauth[cor1]{Corresponding author.}
\address{Physics Department, Yale University, New Haven, CT 06520, USA}

\begin{abstract}
We report the preparation of neutron-activated xenon for the calibration of liquid xenon (LXe) detectors. Gamma rays from the decay of xenon metastable states, produced by fast neutron activation, were detected and their activities measured in a LXe scintillation detector. Following a five-day activation of natural xenon gas with a $\rm^{252}Cf$ ($4 \times 10^5$ n/s) source, the activities of two gamma ray lines at 164 keV and 236 keV, from $\rm^{131m}Xe$ and $\rm^{129m}Xe$ metastable states, were measured at about 95 and 130 Bq/kg, respectively. We also observed three additional lines at 35 keV, 100 keV and 275 keV, which decay away within a few days. No long-lifetime activity was observed after the neutron activation. 
\end{abstract}

\begin{keyword}
Liquid xenon; Neutron activation; Dark matter searches
\PACS 29.40.Mc, 28.20.-v, 95.35.+d
\end{keyword}
\end{frontmatter}

\section{Introduction}
Liquid xenon is increasingly used as a detection material, due to its good ionization and scintillation properties, for applications in $\gamma$-ray calorimetry~\cite{Carugno:NIM96},  $\gamma$-ray Compton imaging telescopes \cite{Aprile:LXeGRIT}, and searches for $\mu \rightarrow e\gamma$~\cite{MEG} and neutrinoless double beta decay~\cite{EXO}. In recent years, LXe has also attracted interest as a material for direct dark matter detection \cite{XENON,Ni:thesis,Yamashita:03,Akimov:2006qw,Alner:2007ja}, in which hypothetical weakly interacting massive particles (WIMPs) can be detected from their elastic scattering from target xenon nuclei, producing a signal with energy of a few keV. LXe satisfies the requirements for a sensitive dark matter detection due to its scalability to large masses, good background discrimination and low-energy threshold. A 15~kg LXe detector (XENON10) was built and deployed in the Gran Sasso National Laboratory and was used to take WIMP-search data in 2006 and early 2007. LXe dark matter detectors with mass on the 100~kg scale were proposed recently~\cite{LUX}. 

These detectors are designed for low-energy (a few keV) event detection and use a two-phase (liquid/gas) technique to realize 3D position sensitivity \cite{ni:ucla06}, which is important for position-dependent corrections of the signals and background rejection through fiducial-volume cuts. Thus precise energy and position calibrations are important for these detectors. However, external low-energy gamma-rays, such as 122 keV gamma rays from $^{57}$Co, will be stopped within a few mm near the detector's edge, due to LXe's high density (about 3 g/cm$\rm^3$) and high stopping power for low-energy gamma rays. High energy gamma rays, such as 662 keV gamma rays from $^{137}$Cs, will make Compton scatters in LXe. Because the expected WIMP signal is at only a few keV, the non-proportionality of LXe scintillation yield at different gamma-ray energies \cite{Yamashita:04,Ni:scint} and non-linearity in the photomultipliers (PMTs) and readout electronics can introduce additional systematic uncertainties when high-energy gamma rays are used for calibration.

One possibility for a reliable and precise detector calibration is to use gamma rays from decay of xenon metastable states, such as $\rm^{129m}Xe$ and $\rm^{131m}Xe$, which emit gamma rays at 236 keV and 164 keV with half-lives of 8.9 and 11.8 days respectively \cite{toi}. They can be produced by thermal-neutron capture on $\rm^{128}Xe$ and $\rm^{130}Xe$, or by fast neutron inelastic scattering on $\rm^{129}Xe$ and $\rm^{131}Xe$. The activities of these metastable states can be built up by placing a fast neutron source (e.g. $^{252}$Cf) around the xenon target. The activated xenon then can be introduced into the dark matter detector as an internal calibration source. The activated xenon will disperse evenly throughout the detector, allowing prompt scintillation light and ionization signals to be quantified as a function of position. The activities will decay away in a few weeks, to satisfy the low-background requirement for a continuing dark matter search. In order not to build up long-lifetime activities in the dark matter detector's construction materials, it is important to prepare the activated xenon far away from the detector itself and introduce the activated xenon into the detector afterwards. In this paper, we report on the preparation and measured activities from neutron-activated xenon.

\section{Activated xenon preparation and experimental setup}
\begin{figure}[htbp]
\begin{center}
\includegraphics[width=0.9\textwidth]{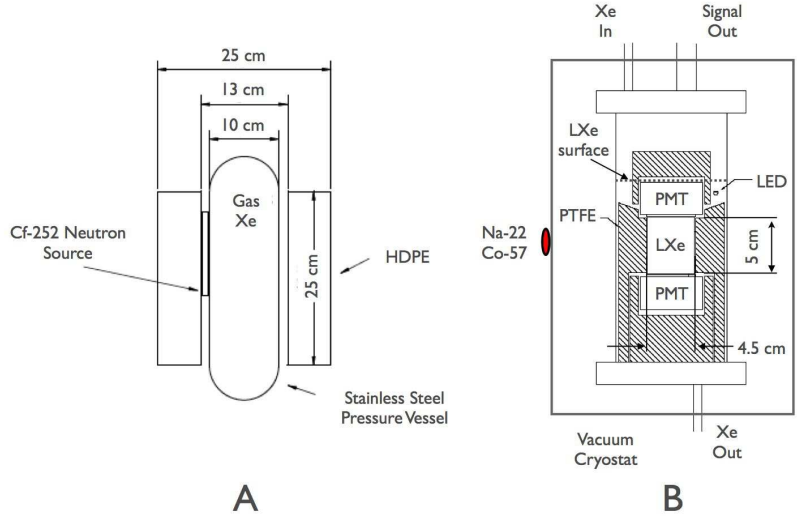}
\caption{A: Setup for xenon activation using a $\rm^{252}Cf$ neutron source. B: A schematic of the LXe scintillation detector for measuring xenon activity. The active LXe target, between the two PMTs, is 4.5 cm in diameter and 5 cm in length (about 240~g in mass).}
\label{CellDrawing}
\end{center}
\end{figure}

To prepare the activated xenon, a high pressure 2.2-liter steel cylinder containing 1~kg of xenon gas was surrounded by a high density polyethylene (HDPE) moderator, as shown in Fig.~\ref{CellDrawing}A. A $\rm^{252}Cf$ ($4 \times 10^5$ n/s) neutron source was placed between the steel cylinder and the HDPE moderator. The HDPE moderator was used to enhance fast neutron interaction with the xenon and also to produce thermal neutrons. The xenon was irradiated continuously by the neutron source for five days. 

The xenon activities are built up by two channels. In the first, fast neutron inelastic scattering produces metastable states. In the other channel, thermal neutrons capture on the Xe nuclei. An estimation, based on the neutron flux, activation period and inelastic or thermal neutron interaction cross-sections in xenon (see Table~\ref{table1}), predicts that the main activity is from $\rm^{129m}Xe$ and $\rm^{131m}Xe$ ($\sim$100 Bq/kg), produced by neutron inelastic scattering on $\rm^{129}Xe$ and $\rm^{131}Xe$. Other possible activity from the activated xenon is summarized in Table~\ref{table1}. 

\begin{sidewaystable}
\centering

\caption{\label{table1}Main activation modes, their association gamma ray energy and a rough estimation of the decay rates.}
\begin{minipage}{\textwidth}
\centering
\begin{tabular}{ccccccccc}
\hline
\hline
Xe & Abundance & 	Reaction		 & Cross-section	& Daughter	&Decay  &	Decay		& $E_\gamma$\footnote{The main gamma ray energy associated with the decay of daughter products} 	& Rate\footnote{Order of magnitude estimation for event rates, one day after the activation ended, based on activation with $10^4$ n/s of thermal neutrons and $10^5$ n/s of fast neutrons passing through the 1~kg Xe target for five days.}    (Bq/kg)  \\
Isotope & (atom \%) & Mode &  (barn)		& Product & Mode	& Half-life	& (keV) & t = 1 day  \\
\hline
$\rm^{124}Xe$	&	0.09	&	$\rm^{124}Xe(n,\gamma)^{125m}Xe$	&	28 &	$\rm^{125m}Xe$&	IT	&	57 s & 252.8 &  $<$1e-6\\
			&		&	$\rm^{124}Xe(n,\gamma)^{125}Xe$		&147	& $\rm^{125}Xe$	 &	$\beta^+$,EC	&	17 hr & 188/243.4 & 100 \\
			&		&	$\rm^{125}Xe\rightarrow^{125}I$		& ---	&$\rm^{125}I$ &		 EC & 59.4 d & 35.5 & 1 \\
$\rm^{129}Xe$	&	26.4	&	$\rm^{129}Xe(n,n')^{129m}Xe$	&	1.6	& $\rm^{129m}Xe$ &	IT	&	8.9 d	& 236.1 & 100 \\
$\rm^{131}Xe$	&	21.2	&	$\rm^{131}Xe(n,n')^{131m}Xe$	&	1.3	& $\rm^{131m}Xe$ &	IT	&	11.8 d & 163.9 & 100\\
$\rm^{132}Xe$	&	26.9	&	$\rm^{132}Xe(n,\gamma)^{133m}Xe$	& 	0.05	& $\rm^{133m}Xe$ &	IT	&	2.2 d & 233.2 & 1\\
			&		&	$\rm^{132}Xe(n,\gamma)^{133}Xe$		& 	0.4	& $\rm^{133}Xe$ &	$\beta^-$	&	5.2 d & 81.0 &  10 \\
$\rm^{136}Xe$	&	8.87	&	$\rm^{136}Xe(n,\gamma)^{137}Xe$		&	0.23& $\rm^{137}Xe$ &	$\beta^-$	&	3.8 m & 455.5 & $<$1e-6\\
			&		&	$\rm^{137}Xe\rightarrow^{137}Cs$			& 	---	& $\rm^{137}Cs$ &	$\beta^-$ &	30.1 y & 661.6 & 1e-3 \\
\hline
\hline
\end{tabular}
\end{minipage}

\end{sidewaystable}


\begin{figure}[htbp]
\begin{center}
\includegraphics[width=0.8\textwidth]{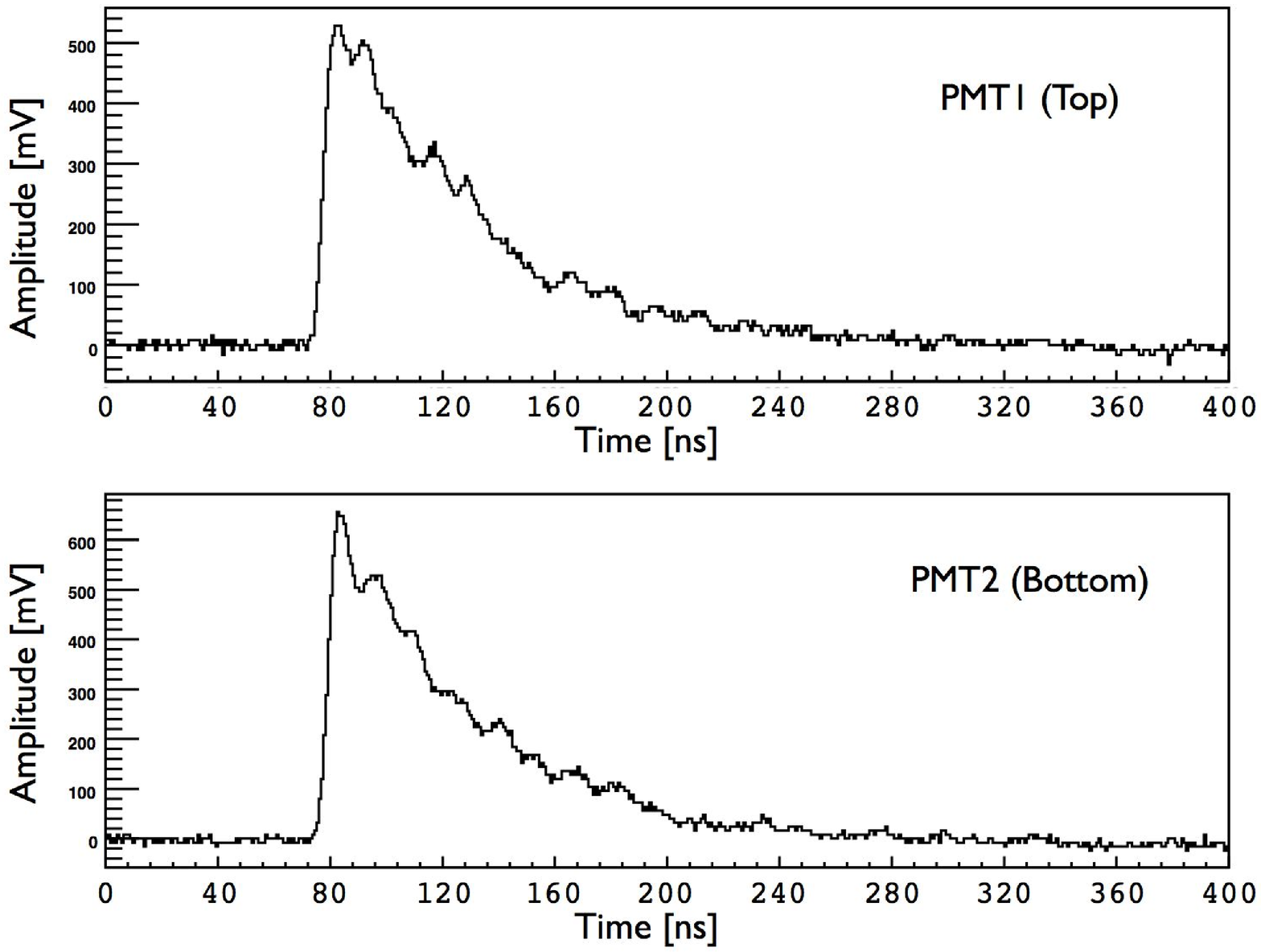}
\caption{Typical scintillation light waveforms for a 122 keV gamma ray event in the scintillation detector. The top PMT (PMT1) detected about 308 photoelectrons (pe) and the bottom one (PMT2) detected 408 pe for this event.}
\label{waveform}
\end{center}
\end{figure}

The activity from neutron-activated xenon was measured with a LXe detector, attached to a two-stage cold finger system and maintained at a stable temperature of 165 K by a pulse-tube refrigerator \cite{PTR}. The activated xenon gas was liquified in a small reservoir attached to the first-stage of the cold finger system, and transferred by gravity into a scintillation detector (Figure~\ref{CellDrawing}B) attached to the second stage of the cold finger system. In the scintillation detector, the activated LXe target is viewed by two VUV sensitive photomultipliers \cite{R9288} immersed in the LXe and is surrounded by a teflon (PTFE) cylinder to enhance the light collection efficiency \cite{Yamashita:04}. To maintain the LXe purity, the xenon is continuously extracted from the bottom of the LXe scintillation detector, vaporized, and circulated through a high-temperature getter purifier \cite{purifier} via a diaphragm pump. After a few hours of recirculation and purification, the scintillation light yield increased to a maximum and became stable, which indicates a good LXe purity.  A blue LED is used for PMT gain calibration. The PMT gains were on the order of $10^6$ with bias voltages of 900 V.

The LXe scintillation light detected by the PMTs are split into two copies by a fan-in/out. One copy goes into NIM electronics for triggering. Another copy goes into a 350~MHz oscilloscope \cite{Tektronix} for data-taking. A coincidence signal within 100 ns between the two PMTs is required for a trigger. A hardware counter is used to record the total number of triggers, which is used to calculate the live-time. The scintillation light waveforms from the PMTs are digitized with a sampling time of 0.8 ns for 400 ns (see Fig. \ref{waveform}). Signals are expressed in number of photoelectrons (pe), obtained by dividing the pulse area by the PMT's single-photoelectron response.

\section{Energy calibration}

The energy is calibrated with external gamma ray sources ($^{57}$Co and $^{22}$Na), placed outside of the vacuum cryostat (see Fig.~\ref{CellDrawing}B). The energy spectra are obtained from the distribution of the summed number of photoelectrons from two PMTs ($N_{pe}^{tot} = N_{pe}^{1} + N_{pe}^2$), where $N_{pe}^1$ and $N_{pe}^2$ are the number of photoelectrons from the two PMTs separately. Based on a Monte Carlo simulation, the light collection efficiency in the detector is strongly dependent on the event location along the Z-axis of the LXe target cylinder. The PMT near the event site sees more light, while the other PMT sees less. Thus an event position can be calculated based on the signal asymmetry, defined as $A = (N_{pe}^1-N_{pe}^2)/(N_{pe}^1+N_{pe}^2)$. In our analyses, we only choose events located in the middle of the scintillation detector, with $|A|<0.2$. The chosen range of the $A$ value avoids the position dependence of the signal and improves the energy resolution of the scintillation light spectra.

\begin{figure}[htbp]
\begin{center}
\includegraphics[width=0.7\textwidth]{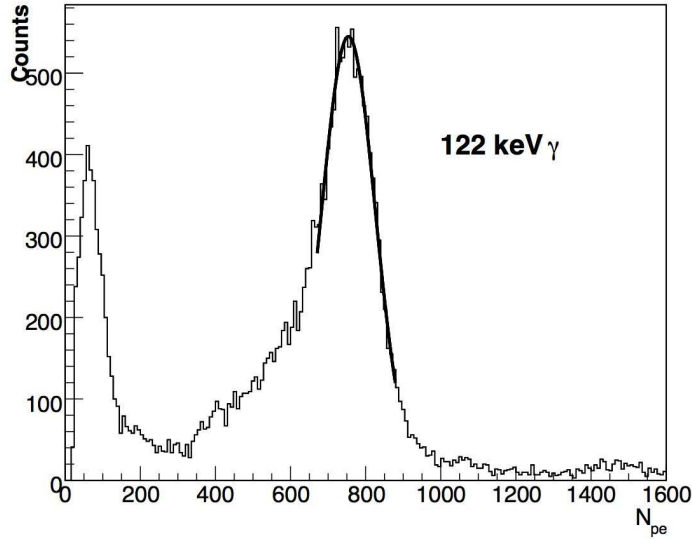}
\caption{Scintillation spectrum for a $\rm^{57}$Co gamma ray source. The shoulder below the 122~keV gamma peak is likely due to a non-uniform light collection efficiency at the edge of the detector. A Gaussian fit of the 122~keV peak gives a light yield of 754 pe and a resolution of 72~pe ($\sigma$), or 9.5\%.}
\label{Co57}
\end{center}
\end{figure}

\begin{figure}[htbp]
\begin{center}
\includegraphics[width=0.7\textwidth]{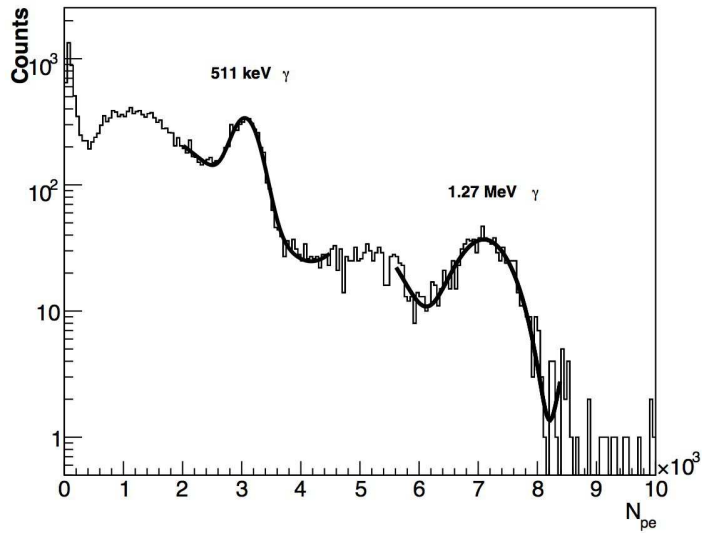}
\caption{Scintillation spectrum for a $\rm^{22}Na$ gamma ray source. The two gamma-ray peaks are fit by a summed Gaussian and a second order polynomial function (solid curves). The 511~keV and 1.27~MeV gamma ray peaks give 3067~pe and 7122~pe, separately. The energy resolutions for these two gamma ray lines are both 7.3\% ($\sigma$).}
\label{Na22}
\end{center}
\end{figure}

Fig. \ref{Co57} shows the scintillation light spectrum from $^{57}$Co source. The 122~keV gamma ray peak gives a total of 754 pe (6.2~pe/keV) and an energy resolution of 9.5\%($\sigma$). Fig.~\ref{Na22} shows the scintillation light spectrum from $^{22}$Na gamma source. The 511~keV and 1.27~MeV gamma ray peaks give 3067~pe and 7122~pe, separately. The resolutions for these two peaks are very similar, at about 7.3\%($\sigma$).

\begin{figure}[htbp]
\begin{center}
\includegraphics[width=0.7\textwidth]{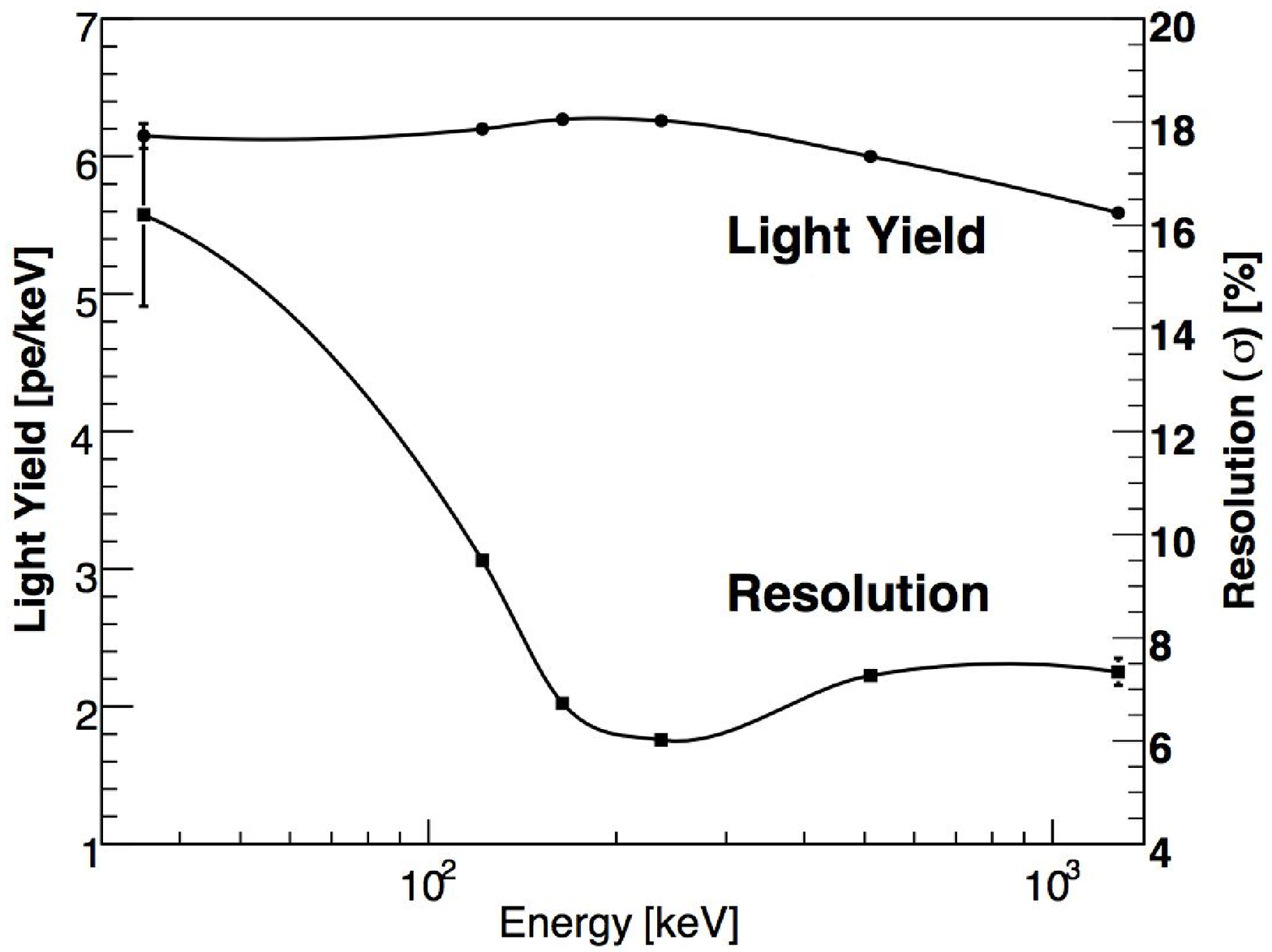}
\caption{Light yields and energy resolutions of scintillation light in LXe for gamma rays at different energies in the current detector.The energy resolutions for energies above 511 keV are slightly worse than at low energy, which can be explained by the smearing due to Compton scattering of the high energy gamma rays. Statistical errors are negligible except for the lowest energy point.}
\label{yield_res}
\end{center}
\end{figure}

Fig.~\ref{yield_res} shows the light yields and energy resolutions for these calibration peaks, as well as the 35 keV X-ray peak and two xenon activation peaks at 164 keV and 236 keV (see next section). The light yields for gamma rays below 236~keV have a very linear response, while the high energy gamma rays apparently give less light yield per unit energy than those from lower energy gamma rays. This is due to the different linear energy transfer (LET) at different gamma ray energies. This detector's capabilities are comparable to those of other recently-demonstrated LXe scintillation detectors \cite{Yamashita:04,Ni:scint,Doke:01}. 

\section{Results from activated xenon}

\begin{figure}[htbp]
\begin{center}
\includegraphics[width=0.95\textwidth]{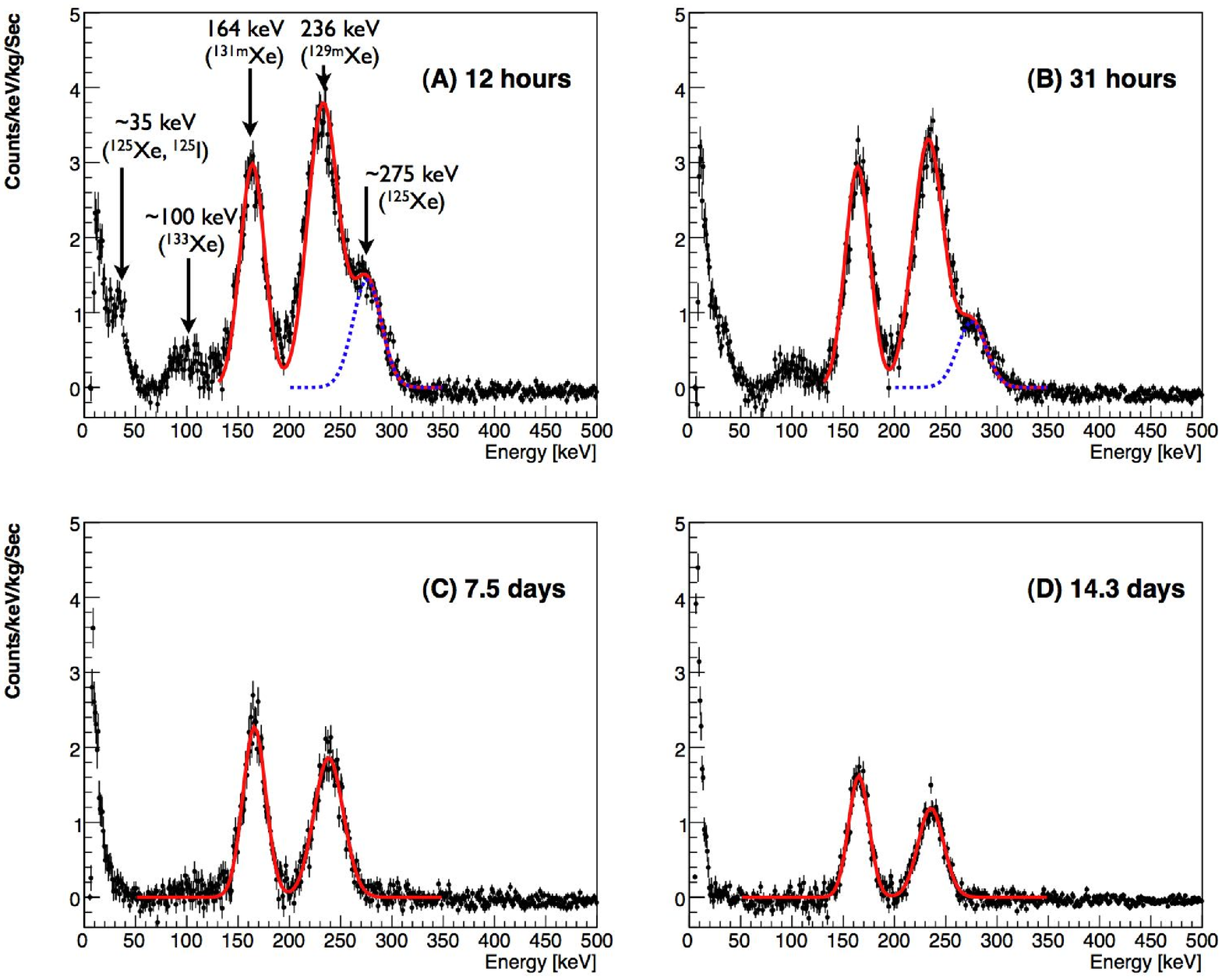}
\caption{Background subtracted spectra from activated xenon, 12 hours (A), 31 hours (B), 7.5 days (C) and 14.3 days (D), after the 5-day neutron-activation. All spectra show clearly the two gamma peaks at 164~keV($\rm^{131m}Xe$) and 236~keV($\rm^{129m}Xe$). The energy resolutions for the 164 and 236~keV peaks are 7.0\%($\sigma$) and 5.1\%, respectively.}
\label{Act_subbkg}
\end{center}
\end{figure}

\begin{figure}[htbp]
\begin{center}
\includegraphics[width=0.8\textwidth]{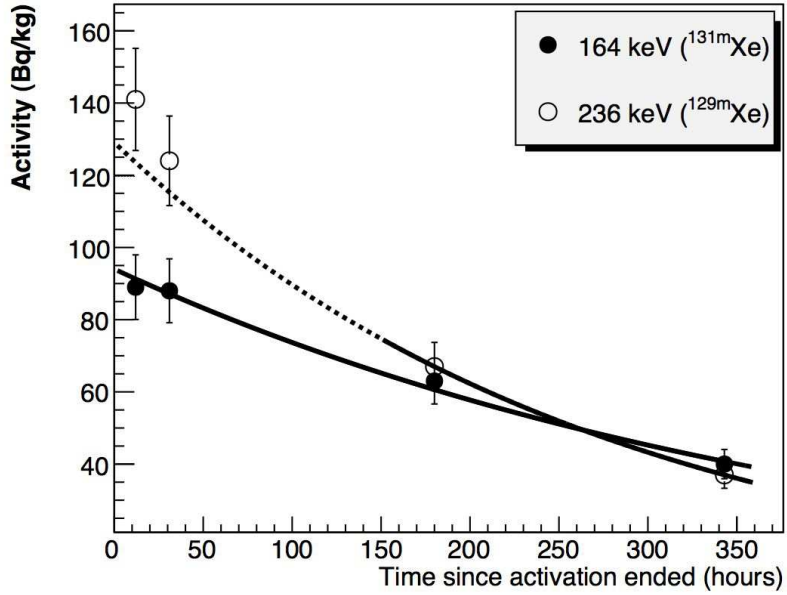}
\caption{Activities from $\rm^{131m}Xe$ and $\rm^{129m}Xe$ at different time after the neutron activation. Exponential functions are used to fit the data points. The half-lives from the fits are found to be $12.2\pm1.8$ days for $\rm^{131m}Xe$ and $8.4\pm2.0$ days for $\rm^{129m}Xe$.}
\label{decay}
\end{center}
\end{figure}

In order to measure the activity from the neutron-activated Xe, we first measured the background activity in our laboratory. The background subtracted spectra, measured at 12 hours, 31 hours, 7.5 days and 14.3 days after the five-day activation ended, are shown in Fig. \ref{Act_subbkg}. We fit the two earlier spectra [Fig. \ref{Act_subbkg} (A) and (B)] with a sum of three Gaussian functions, for the 164~keV, 236~keV and 275~keV peaks. For the two later runs [Fig. \ref{Act_subbkg} (C) and (D)], we fit the spectra with a sum of two Gaussian functions, for the 164~keV and 236~keV peaks. The integrated activities from these peaks are listed in Table~\ref{table2}. The drop of activity around the 236 keV peak between the two early measurements is mainly due to the decay of  $\rm^{133m}Xe$, which emits 233~keV gamma rays with a half-life of 2.2 days. The change in activity with time for the 164~keV and 236~keV lines is plotted in Fig.~\ref{decay} with exponential fits. For the 236 keV gamma peak, only the two points at longer times are used for the fit since the 236~keV gamma rays at the earlier times are mixed with 233 keV gammas from $\rm^{133m}Xe$. The half-lives from the fits are in consistent with the values in \cite{toi}.

\begin{table}
\centering
\caption{\label{table2}Measured activity and identified sources for observed peaks in the activated xenon.}
\centering
\begin{tabular}{cc|cccc}
\hline
\hline
Peak Energy &  Identified  &  &  Rate      &      (Bq/kg)   &     \\ 
 (keV)  &  Source & t = 12 h  & t = 31 h & t = 7.5 d & t = 14.3 d     \\
\hline
$\sim$35   &  $\rm^{125}Xe$, $\rm^{125}I$ &   11  &   3  &   & \\
$\sim$100 & $\rm^{133}Xe$  &   15   &  9   &   & \\
164 &  $\rm^{131m}Xe$ &    90    &   89   &  63   &   40  \\
236 &  $\rm^{129m}Xe$ &     141  &   124   & 67   & 37  \\
$~\sim$275 & $\rm^{125}Xe$ &    51  & 30  &     & \\
\hline
\hline
\end{tabular}
\end{table}

In addition to the gamma peaks at 164 keV ($\rm^{131m}Xe$) and 236 keV ($\rm^{129m}Xe$), the two early measurements, within two days after the neutron-activation, also show clear lines at around 35 keV, 100 keV and 275 keV. The activities of these lines are listed in Table~\ref{table2}. The 100 keV peak is possibly due to $\rm^{133}Xe$, which has a $\beta$-decay followed by a 81~keV gamma ray. The 275~keV peak is likely from $\rm^{125}Xe$ EC decay, which has a 16.9-hour half-life and produces a 243 keV gamma ray coincident with a 29 keV characteristic X-ray from K-shell electrons. The 100~keV and 275~keV peaks found in this study were also previously observed in a neutron-activated high pressure xenon detector \cite{Dmitrenko}.

\begin{figure}[htbp]
\begin{center}
\includegraphics[width=0.8\textwidth]{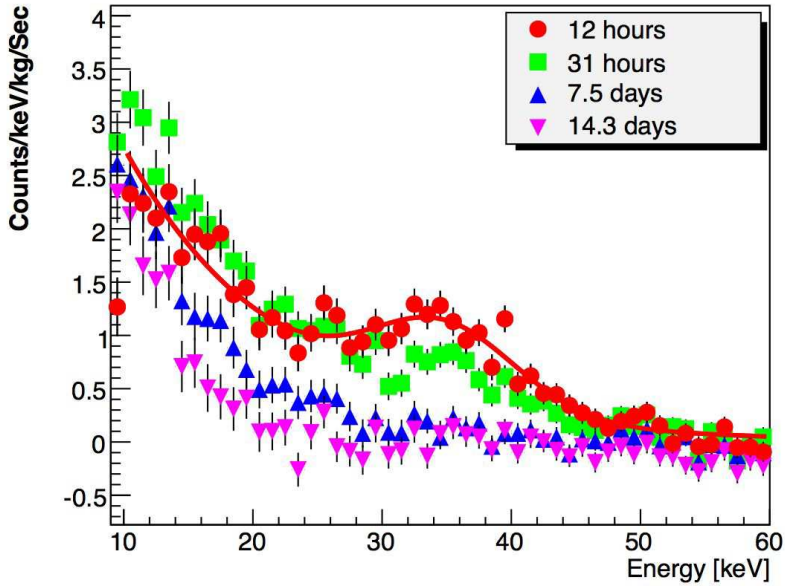}
\caption{Background subtracted spectra for energy below 60 keV from the activated xenon at different time after the activation ended.}
\label{low_energy}
\end{center}
\end{figure}

The low-energy spectra are examined in detail in Fig. \ref{low_energy}. It is clear that the first measurement with the activated xenon (12 hours after the activation ended) shows a peak at around 35 keV. The resolution of the 35 keV peak is about 17\% ($\sigma$). This peak is possibly a combination of the 29 keV characteristic X-ray from the decay of $\rm^{125}Xe$, and the daughter $\rm^{125}I$ decay which emits a 35~keV gamma ray with a 59 day half-life.  The runs one week after activation do not show any clear activity in the 35 keV peak, due to the short half-life of $\rm^{125}Xe$ and the removal of $\rm^{125}I$ by the purification getter.

\section{Conclusion}
A LXe scintillation detector with two PMTs immersed in the liquid has been used to study xenon activity following neutron activation. The high light yield of the scintillation detector allows identification of activities down to 10~keV. After five days of neutron activation of 1~kg xenon gas with a $4\times10^5$ neutron/s $^{252}$Cf source placed near a two-liter gas cylinder and surrounded by a 2.5-inch-thick high density polyethylene moderator, the activities from two gamma ray lines from $\rm^{129m}Xe$ and $\rm^{131m}Xe$ were measured to be around 100~Bq/kg right after the activation ended. We also observed lines at around 35 keV, 100 keV and 275 keV, with initial activities below 100 Bq/kg and half-lives less than a few days. The only measurable activity after two weeks following the neutron-activation are the 164 keV and 236 keV gammas from $\rm^{131m}Xe$ and $\rm^{129m}Xe$ decay. These two gamma peaks are very useful for a precise energy and position calibration of a large LXe detector. The activated-xenon prepared in this study has been successfully used in calibration of the XENON10 dark matter detector at the Gran Sasso Laboratory \cite{Aprile:PRL07}. 

\section{Acknowledgement}
We would like to thank George Andrews and Kevin Charbonneau of the Yale  Radiation Safety for providing the Cf-252 neutron source for this work. This work was supported by National Science Foundation under grant No.~PHY-04-00596.


\end{document}